\documentclass{article}

\usepackage{microtype}
\usepackage{graphicx}
\usepackage{subfigure}
\usepackage{booktabs} 

\usepackage{amsfonts}       
\usepackage{amsmath}

\usepackage{algorithm}
\usepackage{algpseudocode}

\usepackage{hyperref}

\newcommand{\eg}{\textit{e.g.}\ }
\newcommand{\ie}{\textit{i.e.}\ }

\usepackage[accepted]{icml2019}

\icmltitlerunning{} 

\begin{document}

\twocolumn[
\icmltitle{Memory-efficient Learning for Large-scale Computational Imaging \\- NeurIPS deep inverse workshop -}




\begin{icmlauthorlist}
\icmlauthor{Michael Kellman}{one}
\icmlauthor{Jon Tamir}{one}
\icmlauthor{Emrah Bostan}{one}
\icmlauthor{Michael Lustig}{one}
\icmlauthor{Laura Waller}{one}
\end{icmlauthorlist}

\icmlaffiliation{one}{Electrical Engineering and Computer Sciences, University of California, Berkeley, USA}

\icmlcorrespondingauthor{Michael Kellman}{\url{kellman@berkeley.edu}}

\icmlkeywords{}

\vskip 0.3in
]




\section*{Abstract} 
    Computational imaging systems jointly design computation and hardware to retrieve information which is not traditionally accessible with standard imaging systems. Recently, critical aspects such as experimental design and image priors are optimized through deep neural networks formed by the unrolled iterations of classical physics-based reconstructions (termed physics-based networks). However, for real-world large-scale systems, computing gradients via backpropagation restricts learning due to memory limitations of graphical processing units. In this work, we propose a memory-efficient learning procedure that exploits the reversibility of the network's layers to enable data-driven design for large-scale computational imaging. We demonstrate our method’s practicality on two large-scale systems: super-resolution optical microscopy and multi-channel magnetic resonance imaging.


\section{Introduction}
\label{sec:intro}

Computational imaging systems (tomographic systems, computational optics, magnetic resonance imaging, to name a few) jointly design software and hardware to retrieve information which is not traditionally accessible on standard imaging systems. Generally, such systems are characterized by how the information is encoded (forward process) and decoded (inverse problem) from the measurements. The decoding process is typically iterative in nature, alternating between enforcing data consistency and image prior knowledge. Recent work has demonstrated the ability to optimize computational imaging systems by unrolling the iterative decoding process to form a differentiable Physics-based Network (PbN)~\cite{Gregor:2010, sun2016deep, kellman2019physics} and then relying on a dataset and training to learn the system's design parameters, \eg experimental design~\cite{ kellman2019physics, kellman2019data, Sitzmann:2018bd}, image prior model~\cite{Gregor:2010, sun2016deep, aggarwal2018modl, diamond2017unrolled}. PbNs are constructed from the operations of reconstruction, \eg proximal gradient descent algorithm. By including known structures and quantities, such as the forward model, gradient, and proximal updates, PbNs can be efficiently parameterized by only a few learnable variables, thereby enabling an efficient use of training data~\cite{aggarwal2018modl} while still retaining robustness associated with conventional physics-based inverse problems.

Training PbNs relies on gradient-based updates computed using backpropagation (an implementation of reverse-mode differentiation~\cite{Griewank2018evalder}). Most modern imaging systems seek to decode ever-larger growing quantities of information (gigabytes to terabytes) and as this grows, memory required to perform backpropagation is limited by the memory capacity of modern graphical processing units (GPUs).

Methods to save memory during backpropagation (\eg forward recalculation, reverse recalculation, and checkpointing) trade off spatial and temporal complexity~\cite{Griewank2018evalder}. For a PbN with $N$ layers, standard backpropagation achieves $\mathcal{O}(N)$ temporal and spatial complexity. Forward recalculation achieves $\mathcal{O}(1)$ memory complexity, but has to recalculate unstored variables forward from the input of the network when needed, yielding $\mathcal{O}(N^2)$ temporal complexity. Forward checkpointing smoothly trades off temporal, $\mathcal{O}(NK)$, and spatial, $\mathcal{O}(N/K)$, complexity by saving variables every $K$ layers and forward-recalculating unstored variables from the closest checkpoint.

Reverse recalculation provides a practical solution to beat the trade off between spatial vs. temporal complexity by calculating unstored variables in reverse from the output of the network, yielding $\mathcal{O}(N)$ temporal and $\mathcal{O}(1)$ spatial complexities. Recently, several reversibility schemes have been proposed for residual networks~\cite{behrmann2018invertible}, learning ordinary differential equations~\cite{chen2018neural}, and other specialized network architectures~\cite{gomez2017reversible, chang2018reversible}.

In this work, we propose a memory-efficient learning procedure for backpropagation for the PbN formed from proximal gradient descent, thereby enabling learning for many large-scale computational imaging systems. Based on the concept of invertibility and reverse recalculation, we detail how backpropagation can be performed without the need to store intermediate variables for networks composed of gradient and proximal layers. We highlight practical restrictions on the layers and introduce a hybrid scheme that combines our reverse recalculation methods with checkpointing to mitigate numerical error accumulation. Finally, we demonstrate our method's usefulness to learn the design for two practical large-scale computational imaging systems: super-resolution optical microscopy (Fourier Ptychography) and multi-channel magnetic resonance imaging.

\section{Background}
\label{sec:background}
Computational imaging systems are described by how sought information is encoded to and decoded from a set of measurements. The encoding of information, $\mathbf{x}$ into measurements, $\mathbf{y}$, is given by
\begin{align}
    \mathbf{y} = \mathcal{A}(\mathbf{x}) + \mathbf{n},
\end{align}
where $\mathcal{A}$ is the forward model that characterizes the measurement system physics and $\mathbf{n}$ is random system noise. The forward model is a continuous process, but is often approximated by a discrete representation. The retrieval of information from a set of measurements, \ie decoding, is commonly structured using an inverse problem formulation,
\begin{align}
    \mathbf{x}^{\star} = \arg\underset{\mathbf{x}}{\min}\ \mathcal{D}(\mathbf{x};\mathbf{y}) + \mathcal{P}(\mathbf{x}),
    \label{eq:decoder}
\end{align}
where $\mathcal{D}(\cdot)$ is a data fidelity penalty and $\mathcal{P}(\cdot)$ is a prior penalty. When $\mathbf{n}$ is governed by a known noise model, the data consistency penalty can be written as the negative log-likelihood of the appropriate distribution. When $\mathcal{P}(\cdot)$ is a non-smooth prior (\eg $\ell_1$, total variation), proximal gradient descent (PGD) and its accelerated variants are often efficient algorithms to minimize the objective in Eq.~\ref{eq:decoder} and are composed of the following alternating steps:
\begin{align}
    \mathbf{z}^{(k)} &= \mathbf{x}^{(k)} - \alpha \nabla_\mathbf{x}\mathcal{D}(\mathbf{x}^{(k)};\mathbf{y}), \label{eq:grad} \\
    \mathbf{x}^{(k+1)} &= \texttt{prox}_{\mathcal{P}}(\mathbf{z}^{(k)}), \label{eq:prox}
\end{align}
where $\alpha$ is the gradient step size, $\nabla_{\mathbf x}$ is the gradient operator, $\text{prox}_{\mathcal{P}}$ is a proximal function that enforces the prior~\cite{parikh2014proximal}, and $\mathbf{x}^{(k)}$ and $\mathbf{z}^{(k)}$ are intermediate variables for the $k^\mathrm{th}$ iteration.

The structure of the PbN is determined by unrolling $N$ iterations of the optimizer to form the $N$ layers of a network (Eq.~\ref{eq:grad} and Eq.~\ref{eq:prox} form a single layer). Specifically, the input to the network is the initialization of the optimization, $\mathbf{x}^{(0)}$, and the output is the resultant, $\mathbf{x}^{(N)}$. The learnable parameters are optimized using gradient-based methods. Common machine learning toolboxes' (\eg PyTorch, Tensor Flow, Caffe) auto-differentiation functionalities are used to compute gradients for backpropagation. Auto-differentiation accomplishes this by creating a graph composed of the PbN's operations and storing intermediate variables in memory.

\section{Methods}
\label{sec:methods}
Our main contribution is to improve the spatial complexity of backpropagation for PbNs by treating the larger single graph for auto-differentiation as a series of smaller graphs. Specifically, consider a PbN, $\mathcal{F}$, composed of a sequence of layers, 
\begin{align}
    \mathbf{x}^{(k+1)} = \mathcal{F}^{(k)}\left(\mathbf{x}^{(k)}; \theta^{(k)}\right),
\end{align}
where $\mathbf{x}^{(k)}$ and $\mathbf{x}^{(k+1)}$ are the $k^{\text{th}}$ layer input and output, respectively, and $\theta^{(k)}$ are its learnable parameters. When performing reverse-mode differentiation, our method treats a PbN of $N$ layers as $N$ separate smaller graphs, processed one at a time, rather than as a single large graph, thereby saving a factor $N$ in memory. As outlined in Alg.~\ref{alg:meld}, we first recalculate the current layer's input, $\mathbf{x}^{(k-1)}$, from its output, $\mathbf{x}^{(k)}$, using $\mathcal{F}^{(k-1)}_{\text{inverse}}$, and then form one of the smaller graphs by recomputing the output of the layer, $\mathbf{v}^{(k)}$, from the recalculated input. To compute gradients, we then rely on auto-differentiation of each layer's smaller graph to compute the gradient of the loss, $\mathcal{L}$, with respect to $\mathbf{x}^{(k)}$ (denoted $\mathbf{q}^{(k)}$) and $\nabla_{\theta^{(k)}} \mathcal{L}$. The procedure is repeated for all $N$ layers in reverse order.
\begin{algorithm}
    \caption{Memory-efficient learning for physics-based networks}
    \label{alg:meld}
    \begin{algorithmic}[1]
        \Procedure{Memory-efficient Backpropagation}{$\mathbf{x}^{(N)}, \mathbf{q}^{(N)}$}
        \State $k \gets N$
        \For{$k > 0$}
        \State $\mathbf{x}^{(k-1)} \gets \mathcal{F}^{(k-1)}_{\text{inverse}}(\mathbf{x}^{(k)}; \theta^{(k-1)})$ \label{subalg:reverse}
        \State $\mathbf{v}^{(k)} \gets \mathcal{F}^{(k-1)}(\mathbf{x}^{(k-1)}; \theta^{(k-1)})$ \label{subalg:forward}
        \State $\mathbf{q}^{(k-1)} \gets \frac{\partial \mathbf{v}^{(k)}}{\partial \mathbf{x}^{(k-1)}}\mathbf{q}^{(k)}$ \label{subalg:backprop}
        \State $\nabla_{\theta^{(k)}} \mathcal{L} \gets \frac{\partial \mathbf{v}^{(k)}}{\partial \theta^{(k)}}\mathbf{q}^{(k)}$ \label{subalg:gradient}
        \State $k \gets k - 1$
        \EndFor
        \State \textbf{return} $\{\nabla_{\theta^{(k)}} \mathcal{L}\}_{k=0}^{N-1}$
        \EndProcedure
    \end{algorithmic}
\end{algorithm}

In order to perform the reverse-mode differentiation efficiently, we must be able to compute each layer's inverse operation, $\mathcal{F}_{\text{inverse}}^{(k-1)}$. The remainder of this section overviews the procedures to invert gradient and proximal update layers.


\subsection{Inverse of gradient update layer}
\label{ssec:inverse_grad}
A common interpretation of gradient descent is as a forward Euler discretization of a continuous-time ordinary differential equation. As a consequence, the inverse of the gradient step layer (Eq.~\ref{eq:grad}) can be viewed as a backward Euler step,
\begin{align}
    \mathbf{x}^{(k)} &= \mathbf{z}^{(k)} + \alpha \nabla_\mathbf{x}\mathcal{D}(\mathbf{x}^{(k)};\mathbf{y}).
\end{align}
This implicit equation can be solved iteratively via the backward Euler method using the fixed point algorithm (Alg.~\ref{alg:grad_fp}). Convergence is guaranteed if
\begin{align}
    \text{Lip}\left(\alpha \nabla_\mathbf{x}\mathcal{D}(\mathbf{x};\mathbf{y})\right) &< 1, \label{eq:lipschitz}
\end{align}
where $\text{Lip}(\cdot)$ computes the Lipschitz constant of its argument~\cite{banach1922operations}. In the setting when $\mathcal{D}(\mathbf{x};\mathbf{y}) = \|\mathbf{A}\mathbf{x} - \mathbf{y}\|^2$ and $\mathbf{A}$ is linear this can be ensured if $\alpha < \frac{1}{\sigma_{max}(\mathbf{A}^H\mathbf{A})}$, where $\sigma_{max}(\cdot)$ computes the largest singular value of its argument. Finally, as given by Banach Fixed Point Theorem, the fixed point algorithm (Alg.~\ref{alg:grad_fp}) will have an exponential rate of convergence~\cite{banach1922operations}.
\begin{algorithm}
    \caption{Inverse for gradient layer}\label{alg:grad_fp}
    \begin{algorithmic}[1]
        \Procedure{Fixed Point Method}{$\mathbf{z},T$}
        \State $\mathbf{x} \gets \mathbf{z}$
        \For{$t < T$}
        \State $\mathbf{x} \gets \mathbf{z} + \alpha \nabla_\mathbf{x}\mathcal{D}(\mathbf{x};\mathbf{y})$
        \State $t \gets t + 1$
        \EndFor
        \State \textbf{return} $\mathbf{x}$
        \EndProcedure
    \end{algorithmic}
\end{algorithm}

\subsection{Inverse of proximal update layer}
The proximal update (Eq.~\ref{eq:prox}) is defined by the following optimization problem~\cite{parikh2014proximal}:
\begin{align}
    \text{prox}_{\mathcal{P}}(\mathbf{z}^{(k)}) = \arg\underset{\mathbf{v}}{\min}\ \frac{1}{2}\|\mathbf{v} - \mathbf{z}^{(k)}\|_2^2 + \mathcal{P}(\mathbf{v}).
\end{align}
For differentiable $\mathcal{P}(\cdot)$, the optimum of which is,
\begin{align}
    \mathbf{x}^{(k+1)} = \mathbf{z}^{(k)} - \nabla_\mathbf{x} \mathcal{P}(\mathbf{x}^{(k+1)}).
\end{align}
In contrast to the gradient update layer, the proximal update layer can be thought of as a backward Euler step~\cite{parikh2014proximal}. This allows its inverse to be expressed as a forward Euler step,
\begin{align}
     \mathbf{z}^{(k)} = \mathbf{x}^{(k+1)} + \nabla_\mathbf{x} \mathcal{P}(\mathbf{x}^{(k+1)}),
\end{align}
when the proximal function is bijective (\eg $\text{prox}_{\ell_2}$). If the proximal function is not bijective (\eg $\text{prox}_{\ell_1}$) the inversion is not straight forward. However, in many cases it is possible to substitute it with a bijective function with similar behavior.

\begin{figure*}[t]
    \centering
    \includegraphics[width=15.5cm]{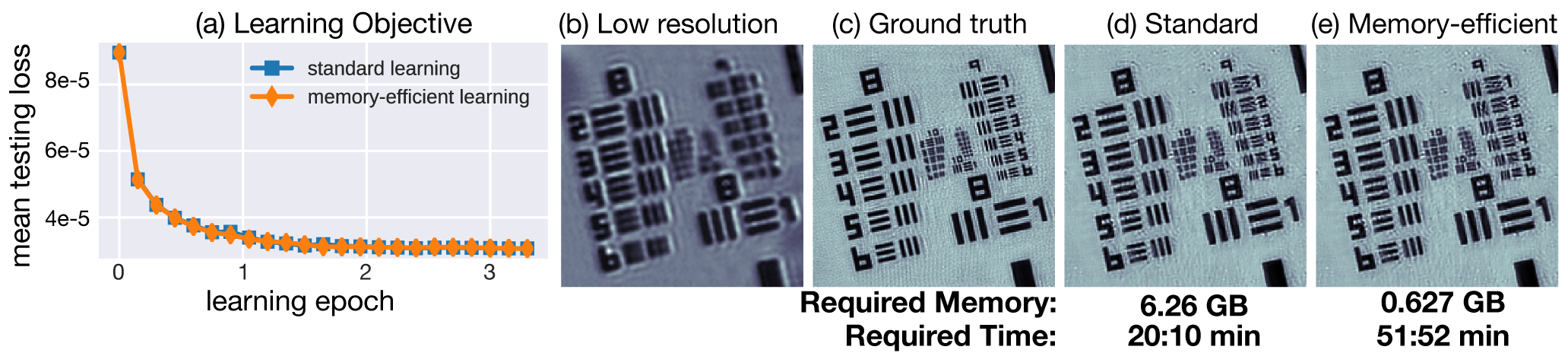}
    \caption{Super-resolution Microscopy: Comparison between (a) mean testing loss for standard and memory-efficient learning techniques. Visualization of (b) low-resolution, (c) ground truth reconstruction using all (89) measurements, and reconstruction using 8 measurements learned using (d) standard (with $6.26$ GB and $20$:$10$ min) and (e) memory-efficient learning (with $0.627$ GB and $51$:$52$ min).}
    \label{fig:fpm_results}
\end{figure*}

\begin{figure*}[t]
    \centering
    \includegraphics[width=15.5cm]{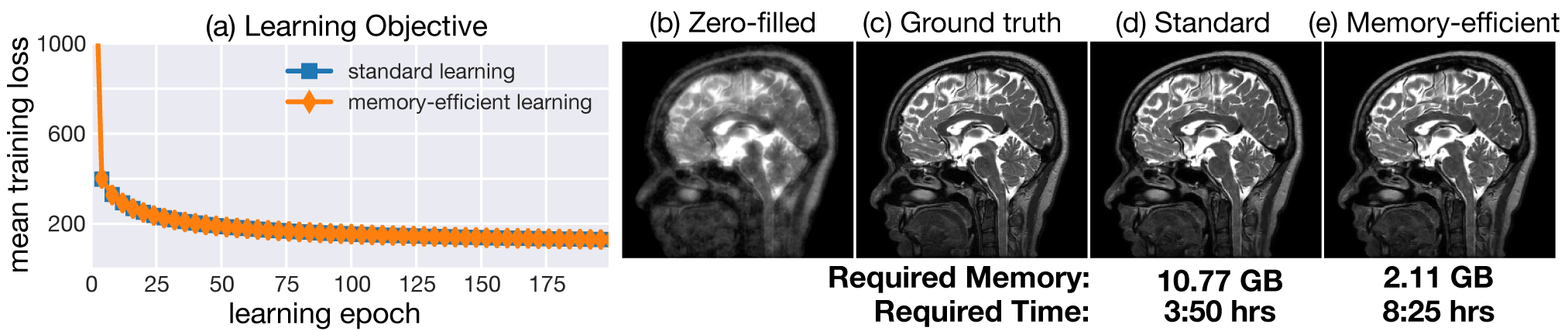}
    \caption{Multi-channel MRI: Comparison between (a) mean training loss for standard and memory-efficient learning techniques. Visualization of (b) zero-filled reconstruction, (c) ground truth reconstruction using fully sampled measurements, and PbN reconstruction learned using (d) standard (with $10.77$ GB and $3$:$50$ hours) and (e) memory-efficient learning (with $2.11$ GB and $8$:$25$ hours).}
    \label{fig:mri_learning}
\end{figure*}
    
\section{Hybrid Reverse Recalculation and Checkpointing}
\label{sec:hybrid}

Reverse recalculation of the unstored variables is non-exact as the operations to calculate the variables are not identical to forward calculation. The result is numerical error between the original forward and reverse calculated variables and as more iterations are unrolled, numerical error can accumulate.

To mitigate these effects, some of the intermediate variables can be stored from forward calculation, referred to as checkpoints. Memory permitting, as many checkpoints should be stored as possible to ensure accuracy while performing reverse recalculation. While most PbNs cannot afford to store all variables required for reverse-mode differentiation, it is often possible to store a few.

\section{Results}
\label{sec:results}

\subsection{Learned experimental design for super resolution optical microscopy}
\label{ssec:fpm}
Standard bright-field microscopy offers a versatile system to image \textit{in vitro} biological samples, however, is restricted to imaging either a large field of view or a high resolution. Fourier Ptychographic Microscopy (FPM)~\cite{Zheng:2013gq} is a super resolution (SR) method that can create gigapixel-scale images beating this trade off on a standard optical microscope by acquiring a series of measurements (up to hundreds) under various illumination settings on an LED array microscopy~\cite{phillips2017quasi} and combining them via a phase retrieval based optimization. The system's dependence on many measurements inhibits its ability to image live fast-moving biology. Reducing the number of measurements is possible using linear multiplexing~\cite{Tian:2015er} and state of the art performance is achieved by forming a PbN and learning its experimental design~\cite{kellman2019data, kellman2019physics}, however, is currently limited in scale due to GPU memory constraints (terabyte-scale memory is required for learning the full measurement system). With our proposed memory-efficient learning framework, we reduce the required memory to only a few gigabytes, thereby enabling the use of consumer-grade GPU hardware.

To evaluate accuracy we compare standard learning with our proposed memory-efficient learning on a problem that fits in standard GPU memory. We reproduce results in \cite{kellman2019data} where the number of measurements are reduced by a factor of $10$ using $6.26$GB of memory using only $0.627$GB and time is only increased by a factor of $2$. To perform memory-efficient learning, we set $T=4$ and checkpoint every $10$ unrolled iterations. The testing loss between our method and standard learning are comparable (Fig.~\ref{fig:fpm_results}a). In addition, we qualitatively highlight equivalence of the two methods, displaying SR reconstructions with learned design using standard (Fig.~\ref{fig:fpm_results}d) and memory-efficient (Fig.~\ref{fig:fpm_results}e) methods. For relative comparison, we display a single low resolution measurement (Fig.~\ref{fig:fpm_results}b) and the ground truth SR reconstruction using all measurements (Fig.~\ref{fig:fpm_results}c).

\subsection{Learned priors for multi-channel MRI}
\label{ssec:mri}
MRI is a powerful Fourier-based medical imaging modality that non-invasively captures rich biophysical information without ionizing radiation. Since MRI acquisition time is directly proportional to the number of acquired measurements, reducing measurements leads to immediate impact on patient throughput and enables capturing fast-changing physiological dynamics. Multi-channel MRI is the standard of care in clinical systems and uses multiple receive coils distributed around the body to acquire measurements in parallel, thereby reducing the total number of required acquisition frames for decoding~\cite{bib:pruessmann1999ys}. By additionally modifying the measurement pattern to take advantage of image prior knowledge, \eg through compressed sensing~\cite{bib:lustig2007hl}, it is possible to dramatically reduce scan times.
As with experimental design, PbNs with learned deep image priors have demonstrated state-of-the-art performance for multi-channel MRI~\cite{Hammernik:2017ku, aggarwal2018modl}, but are limited in network size and number of unrolled iterations due to memory required for training.
Our memory-efficient learning reduces memory footprint at training time, thereby enabling learning for larger problems.

To evaluate our proposed memory-efficient learning, we reproduce the results in~\cite{aggarwal2018modl} for the ``SD-ET-WD'' PbN, which is equivalent to PGD (10 unrolled iterations) where the proximal update is replaced with a learned invertible residual convolutional neural network (RCNN)~\cite{he2016deep,gomez2017reversible,behrmann2018invertible}. We compare training with full backpropagation, requiring $10.77$GB of memory and $3$:$50$ hours, versus memory-efficient learning, requiring $2.11$GB and $8$:$25$ hours. We set $T=6$ and do not use checkpointing. As Fig.~\ref{fig:mri_learning} shows, the training loss is comparable across epochs, and inference results are similar on one image in the training set, with normalized root mean-squared error of 0.03 between conventional and memory-efficient learning.

\section{Remarks}
\label{sec:remarks}
\textbf{Discussion:} Our proposed memory-efficient learning opens the door to applications that are not otherwise possible to train due to GPU memory constraints, without a large increase in training time. While we specialized the procedure to PGD networks, similar approaches can be taken to invert other PbNs with more complex subroutines such as solving linear systems of equations. However, sufficient conditions for invertibility must be met. This limitation is clear in the case of a gradient descent block with an evolving step size, as the Lipschitz constant may no longer satisfy Eq.~\ref{eq:lipschitz}. Furthermore, the convergent behavior of optimization to minima makes accurate reverse recalculation of unstored variables severely ill-posed and can cause numerical error accumulation. Checkpoints can be used to improve the accuracy of reverse recalculated variables, though most PbN are not deep enough for numerical convergence to occur.

\textbf{Conclusion:} In this communication, we presented a practical memory-efficient learning method for large-scale computational imaging problems without dramatically increasing training time. Using the concept of reversibility, we implemented reverse-mode differentiation with favorable spatial and temporal complexities. We demonstrated our method on two representative applications: SR optical microscopy and multi-channel MRI. We expect other computational imaging systems to nicely fall within our framework.

\bibliographystyle{icml2019}
\bibliography{bibtex}

\end{document}